\pdfoutput=1
\documentclass[11pt,twoside,a4paper,cmspaper,final,collab]{cms-tdr}
\def\svnVersion{72741a6}\def\svnDate{2022/06/16}\def\cmsCernNoTag{CERN-EP-2021-270}\def\cmsCernDate{\today}\def\cmsMessage{Published in Physical Review Letters as \href{http://dx.doi.org/10.1103/PhysRevLett.129.021802}{\doi{10.1103/PhysRevLett.129.021802}.}}
\begin{document}\cmsNoteHeader{B2G-20-001}

\newlength\cmsFigWidth
\ifthenelse{\boolean{cms@external}}{\setlength\cmsFigWidth{0.49\textwidth}}{\setlength\cmsFigWidth{0.65\textwidth}}

\newcommand{\dPW}{\ensuremath{\text{deep-}\ifthenelse{\boolean{cms@external}}{\!}{}\PW}\xspace}
\newcommand{\dPWH}{\ensuremath{\text{deep-}\ifthenelse{\boolean{cms@external}}{\!\PW\!\PH}{\PW\PH}}\xspace}
\ifthenelse{\boolean{cms@external}}{\newcommand{\dWorWH}{\ensuremath{\text{deep-}\!\PW\,(\text{deep-}\!\PW\PH)}\xspace}}{\newcommand{\dWorWH}{\ensuremath{\text{deep-}\PW\,(\text{deep-}\PW\PH)}\xspace}}
\newcommand{\WKK}{\ensuremath{\PW_{\text{KK}}}\xspace}
\newcommand{\ZKK}{\ensuremath{\PZ_{\text{KK}}}\xspace}
\newcommand{\MPT}{\ptmiss}
\newcommand{\ptj}{\ensuremath{\pt^\text{j}}\xspace}
\newcommand{\MJ}{\ensuremath{m_{\text{j}}}\xspace}
\newcommand{\MJmax}{\ensuremath{m^{\text{max}}_{\text{j}}}\xspace}
\newcommand{\MJmin}{\ensuremath{m^{\text{min}}_{\text{j}}}\xspace}
\newcommand{\MJlv}{\ensuremath{{m}_{\text{j}\Pell\PGn}}\xspace}
\newcommand{\MJJlv}{\ensuremath{{m}_{\text{jj}\Pell\PGn}}\xspace}
\DeclareRobustCommand{\PR}{{\HepParticle{R}{}{}}\xspace}
\newcommand{\MR}{\ensuremath{m_{\PR}}\xspace}
\newcommand{\MWKK}{\ensuremath{m_{\WKK}}\xspace}
\newcommand{\Rfourq}{\ensuremath{\PR^{4\PQq}}\xspace}
\newcommand{\Rthreeq}{\ensuremath{\PR^{3\PQq}}\xspace}
\newcommand{\Rlqq}{\ensuremath{\PR^{\Pell\PQq\PQq}}\xspace}
\newcommand{\Wjets}{\PW{}\,+\,jets\xspace}

\cmsNoteHeader{B2G-20-001}

\title{Search for resonances decaying to three \texorpdfstring{\PW}{W} bosons in proton-proton collisions at \texorpdfstring{$\sqrt{s} = 13\TeV$}{sqrt(s) = 13 TeV} }

\date{\today}

\abstract{
A search for resonances decaying into a \PW boson and a radion, where the radion decays into two \PW bosons, is presented.  The data analyzed correspond to an integrated luminosity of 138\fbinv recorded in proton-proton collisions with the CMS detector at $\sqrt{s}=13\TeV$.  One isolated charged lepton is required, together with missing transverse momentum and one or two massive large-radius jets, containing the decay products of either two or one \PW bosons, respectively.  No excess over the background estimation is observed. The results are combined with those from a complementary channel with an all-hadronic final state, described in an accompanying paper.  Limits are set on parameters of an extended warped extra-dimensional model.  These searches are the first of their kind at the LHC.
}

\hypersetup{
pdfauthor={CMS Collaboration},
pdftitle={Search for resonances decaying to three W bosons in proton-proton collisions at sqrt(s) = 13 TeV},
pdfsubject={CMS},
pdfkeywords={CMS, triboson}}

\maketitle

The existence of heavy resonances accessible at the CERN LHC is suggested by various theoretical models that try to address limitations of the standard model (SM).
Searching for these resonances in decays to boson pairs (dibosons) has received much attention in recent years~\cite{Aaboud:2018bun,Sirunyan:2019vgt,ATLAS:2020fry,CMS:2021klu,CMS-PAPERS-B2G-18-002,ATLAS:2019nat}.
In the context of such searches, merged jet reconstruction and classification techniques that aim to identify the origin of a large-radius jet from a Lorentz-boosted hadronically decaying particle have been developed and exploited extensively~\cite{Asquith:2018igt}.
Nonetheless, a direct search for a resonance decaying to three vector bosons, a triboson resonance, has never been performed at the TeV scale. 
In the sub-TeV range, a search for a heavy neutral Higgs boson in the $\PZ\PW\PW$ channel has recently been published~\cite{ATLAS:2020gxx}.
The TeV-scale diboson resonance searches~\cite{CMS:2021klu,ATLAS:2020fry,CMS-PAPERS-B2G-18-002,ATLAS:2019nat} are potentially sensitive to a triboson signal. However, as only two of the three bosons are reconstructed, they have not been interpreted in this way.
A search at the TeV-scale is motivated by various theoretical scenarios including extended warped extra-dimensional models presented in 
Refs.~\cite{Aguilar-Saavedra:2015iew,Aguilar-Saavedra:2015rna,Agashe:2016rle,Agashe:2016kfr,Agashe:2017wss,Aguilar-Saavedra:2017iso,Agashe:2018leo,Kuang:2015gka,Ren:2014bza} indicating a discovery potential within LHC reach.
These models provide extensions of the SM that simultaneously address the problems of the Planck--electroweak hierarchy and the origins of flavor structure.

In this Letter and in an accompanying paper~\cite{B2G-21-002}, we present the first searches for massive resonances decaying to three \PW bosons in cascade through $\WKK \to \PW\PR$ and $\PR \to \PW\PW$.
The \WKK is a Kaluza--Klein (KK)~\cite{Randall:1999ee,Randall:1999vf,Agashe:2007zd,Fitzpatrick:2007qr} excited massive gauge boson and \PR is a scalar radion~\cite{Goldberger_1999}. 
The \WKK and \PR bosons are postulated in the Randall--Sundrum extra-dimension scenario~\cite{Randall:1999ee,Randall:1999vf}. The size of the extra dimension is stabilized by introducing a potential with a modulus field~\cite{Randall:1999vf}, resulting in a bulk scalar boson, the radion.

\begin{figure}[h!]\centering
\includegraphics[width=\cmsFigWidth]{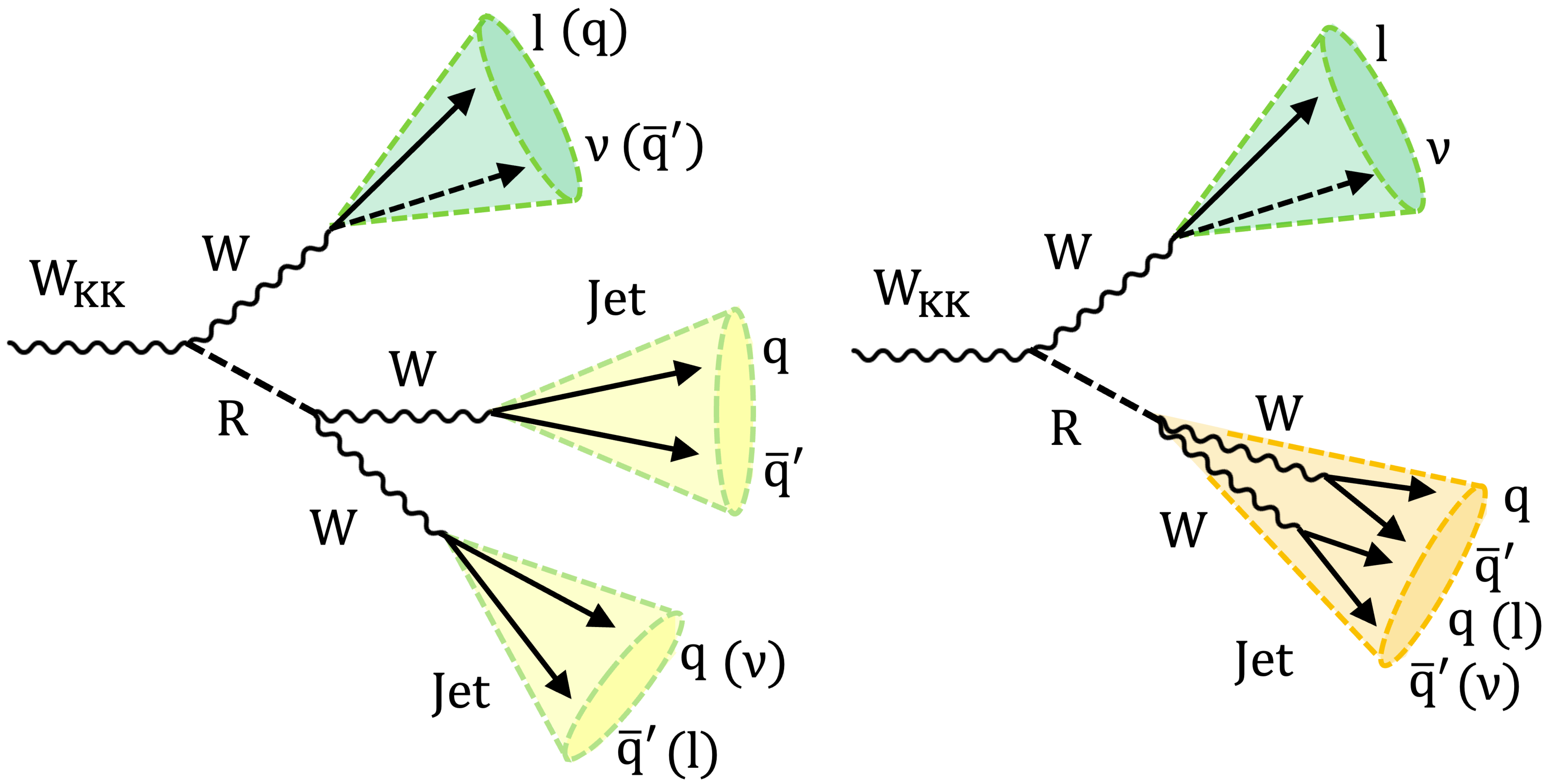}
\caption{Schematic representation of the decay of a KK excitation \WKK to the final states considered in this analysis. 
Left: three individually reconstructed \PW bosons with resolved \PR; right: one individually reconstructed \PW boson and two merged \PW bosons reconstructed as a single large-radius jet.}
\label{fig:diagram}
\end{figure}

We concentrate on the final-state topology comprising one isolated, charged lepton (\Pell), either electron (\Pe) or muon (\Pgm), missing transverse momentum (\MPT), and one or two massive large-radius jets.
A similar topology without an isolated \Pell in the final state is considered in Ref.~\cite{B2G-21-002}.
These two searches use common techniques for jet identification and calibration, which are detailed in Ref.~\cite{B2G-21-002}, while the combination of the two results is presented in this Letter.
The topology studied in this Letter originates from a \PW boson decaying to an isolated \Pell and its neutrino \PGn, and two other \PW bosons decaying into quarks forming hadrons,
which are either reconstructed as two individual merged \PW boson jets, as shown in Fig.~\ref{fig:diagram} (left),
or---depending on the relative masses of the \WKK and \PR resonances---as a single jet containing the decay products of both \PW bosons, as shown in Fig.~\ref{fig:diagram} (right).
We also consider the case where one of the two merged \PW bosons originating from the radion decays leptonically, yielding a nonisolated \Pell inside the jet in addition to the isolated one from the separated \PW boson decay.
The main backgrounds in this analysis are from \Wjets and top quark-antiquark pair (\ttbar) production.
They are estimated using control regions (CRs) with kinematic properties similar to the corresponding signal regions (SRs).
While the analysis is interpreted in terms of one specific model, the search is generic as it is sensitive to many resonant diboson and triboson signals.
For example, resonances decaying into $\PW\PW$ and $\PW\PZ$ can also be detected through this search, although with a lower efficiency than in the dedicated analyses~\cite{CMS:2021klu,ATLAS:2020fry}.
Tabulated results are provided in the HEPData record for this analysis~\cite{hepdata}.

The analysis is based on proton-proton ($\Pp\Pp$) collision data at $\sqrt{s}=13\TeV$ collected by the CMS experiment at the LHC during 2016--2018, corresponding to an integrated luminosity of 138\fbinv~\cite{CMS-LUM-17-003,CMS-PAS-LUM-17-004,CMS-PAS-LUM-18-002}.

The CMS apparatus~\cite{CMS:2008xjf} is a multipurpose, nearly hermetic detector, designed to trigger on~\cite{CMS:2020cmk,CMS:2016ngn} and identify electrons, muons, photons, and (charged and neutral) hadrons~\cite{CMS:2015xaf,CMS:2018rym,CMS:2015myp,CMS:2014pgm}.
A global reconstruction ``particle-flow'' (PF) algorithm~\cite{CMS:2017yfk} combines the information provided by the all-silicon inner tracker and by the crystal electromagnetic and brass-scintillator hadron calorimeters,
operating inside a 3.8\unit{T} superconducting solenoid, with data from gas-ionization muon detectors interleaved with the solenoid return yoke, to build \PGt leptons, jets, \MPT, and other physics objects~\cite{CMS:2018jrd,CMS:2016lmd,CMS:2019ctu}.

Signal events are simulated at leading order (LO) using \MGvATNLO v2.4.2~\cite{Alwall:2014hca} with the recommended parameters according to Refs.~\cite{Agashe:2016rle,Agashe:2016kfr,Agashe:2017wss,Agashe:2018leo}, \ie, the KK gravity coupling $g_{\text{grav}}=6$, the KK gauge couplings $g_{\WKK}=3$ and $g_{\ZKK}=6.708$, and the confinement parameter $\epsilon=0.5$.
In the two-dimensional parameter space, \WKK masses \MWKK from 1.5 to 5.0\TeV and \PR masses \MR from 6 to 90\% of \MWKK are covered.
The decay branching fraction of $\WKK \to \PW\PR \to \PW\PW\PW$ for these parameters typically exceeds 50\%~\cite{Agashe:2017wss}.
For the background simulation, \ttbar production is modeled at next-to-LO (NLO) with \POWHEG v2~\cite{Alioli:2011as}. Quantum chromodynamics multijet production and \Wjets production are simulated at LO with \MGvATNLO.
The other backgrounds are generated at NLO with \MGvATNLO ($\PW\PW$, $s$-channel single \PQt) and \POWHEG ($\PW\PZ$, $\PZ\PZ$, $t$-channel single \PQt, $\PW\PQt$).

The generated events are interfaced with \PYTHIA 8.230~\cite{Sjostrand:2014zea} to simulate the fragmentation, parton shower, and hadronization of partons in the initial and final states, along with the underlying event.
The same simulation settings as for Ref.~\cite{B2G-21-002}, where further details can be found, have been used.
The interactions of all final-state particles with the CMS detector are simulated using \GEANTfour~\cite{Agostinelli:2002hh}.
Simulated events include the contribution of particles from additional $\Pp\Pp$ interactions within the same or nearby bunch crossings (pileup) and are corrected to reproduce the distribution of the number of pileup interactions observed in data.

The events are collected with single-electron or single-muon triggers~\cite{CMS:2020cmk,CMS:2016ngn} and then undergo global event reconstruction based on the PF algorithm~\cite{CMS:2017yfk}.
The PF candidates are corrected for the effect of pileup~\cite{Sirunyan:2020foa}, and are clustered into jets with the anti-\kt algorithm~\cite{Cacciari:2008gp} as implemented in the \textsc{FastJet} package~\cite{Cacciari:2011ma}.
Two distance parameters are used: 0.4 for AK4 jets and 0.8 for large-radius AK8 jets.
The AK4 jets are required to be well separated from any selected AK8 jet with $\Delta R=\sqrt{\smash[b]{(\Delta \eta)^2 + (\Delta \phi)^2}}>0.8$, where $\eta$ is the pseudorapidity and $\phi$ the azimuthal angle.
The quantity \MPT is defined as the magnitude of the vector transverse momentum (\pt) sum of all reconstructed PF candidates in an event.

The AK4 jets arising from \PQb quark hadronization and decay (\PQb jets) are identified using the deep neural network (DNN) algorithm \textsc{DeepCSV},
which takes as input tracks that are displaced from the primary vertex, secondary vertices, and jet kinematic variables~\cite{Sirunyan:2017ezt}.
A working point on the output of the \textsc{DeepCSV} algorithm is chosen such that the efficiency of identifying a \PQb jet is about 65--75\%, while the probability of misidentifying a light-flavor (\Pq) or gluon (\Pg) jet as a \PQb jet is about 1\%.
To identify massive AK8 jets, a ``modified mass-drop'' correction algorithm~\cite{Dasgupta:2013ihk,Butterworth:2008iy},
known as the ``soft-drop'' algorithm~\cite{Larkoski:2014wba} (with parameters  $\beta=0$ and $z_{\text{cut}}=0.1$),
is applied to remove soft and wide-angle radiation from the jet, and the resulting ``groomed jet mass'' (\MJ) is used.

Events with exactly one isolated \Pe (\Pgm) with $\pt>55\GeV$ and $\abs{\eta_{\Pe(\Pgm)}}<2.5\,(2.4)$ and no second isolated \Pe (\Pgm) with $\pt>35\,(20)\GeV$ are selected.
Furthermore, we require $\MPT>80\,(40)\GeV$ for the \Pe (\Pgm) channel.
The \pt of the reconstructed leptonically decaying \PW boson candidate must exceed 200\GeV.
The neutrino is reconstructed as in Ref.~\cite{Sirunyan:2018iff} using \MPT and requiring the effective mass of the $\Pell\PGn$ system to be consistent with the \PW boson mass.
Jets overlapping with the selected isolated lepton within $\Delta R_{\mathrm{j}\Pell}<1.0$ are removed.
Selected events need to have one or two AK8 jets with $\ptj>200\GeV$ and $\abs{\eta^\text{j}}<2.4$.
For events with only one jet, \MJ is required to be greater than 60\GeV, while for events with two jets the jet with maximum \MJ is required to have $60<\MJmax<100\GeV$ and the other jet mass is referred to as $\MJmin$.
Events with identified \PQb jets or more than two AK4 jets are vetoed.
The scalar \pt sum of the reconstructed leptonically decaying \PW boson and the selected AK8 jets is required to be greater than 1\TeV.
The invariant mass of the reconstructed $\Pell\PGn$+jet(s) system, $\MJlv$ or $\MJJlv$ for one or two selected AK8 jets, respectively, must exceed 1.1\TeV.

Several different radion decay topologies are considered.
A collimated radion decay into two merged hadronic \PW boson jets ($\PR \to \PW\PW \to 4\Pq$) yields a single jet containing the decay products of either all four quarks (designated as \Rfourq) or only three of them (\Rthreeq).
A merged radion decay with one of the \PW bosons decaying leptonically ($\PR \to \PW\PW \to \Pell\PGn\Pq\Pq$) yields a jet containing the decay products of two quarks from the $\PW \to \Pq \Paq^{\prime}$ decay as well as an overlapping nonisolated charged lepton. This topology is designated as \Rlqq.
In rare cases where the decay $\PR \to \PW\PW \to \Pell\PGn\Pq\Pq$ results in a lepton that still fulfills the isolation criteria even though it is overlapping with a jet, we remove the overlapping jet and consider the only remaining jet to correspond to the \PW boson from the \WKK resonance decay.
Possibilities other than these contribute less than 5\% of the signal yield and therefore are not considered.

To increase discrimination of signal from background, the substructure of the selected AK8 jets is analyzed using the DNN-based \textsc{DeepAK8} jet classification algorithm~\cite{Sirunyan:2020lcu}.
This algorithm has been trained using simulated events to identify hadronic decays of \PW and Higgs bosons (\PH, in the 4\Pq mode), as well as top quarks,
based on the reconstructed particles and secondary vertices associated with the corresponding jet.
In the default training of the algorithm, the masses of the signal jets are used, and therefore signals with masses different to the ones mentioned above cannot be identified.
Thus, we make use of the algorithm's mass-decorrelated version to identify jets exhibiting substructure compatible with a merged radion decay (\Rfourq, \Rthreeq, \Rlqq), but with arbitrary mass.
For the identification of merged radion jet candidates, we combine the algorithm's outputs to simultaneously discriminate $\PW \to \Pq \Paq^{\prime}$ jets and signal jets similar to $\PH \to \PW\PW \to 4\Pq$ from jets originating from the hadronization of a \Pq or \Pg.
We call the resulting discriminants for merged radion decays ``\dPWH'', and for \PW bosons ``\dPW''.
Similarly, a discriminant named ``deep-\PQt'' is formed to distinguish top quarks from \Pq/\Pg jets.
The \textsc{DeepAK8} discriminant values peak towards unity for the selected type of jets and towards zero for the rejected \Pq/\Pg background jets.
A detailed description of these variables together with their performance for different jet types can be found in Ref.~\cite{B2G-21-002}.

Using the jet mass and the \dPWH and \dPW discriminants, selected events are split into six SRs based on the signal topology.
Jets with $\MJ>100\GeV$ ($60<\MJ<100\GeV$) are considered as radion (\PW boson) candidates and thus required to pass a \dPWH (\dPW) selection, while for lower-mass jets with $\MJ<60\GeV$ no such condition is applied.
For events with one selected jet, targeting the merged radion jet topology, three regions (SR1--3) are defined using different \MJ windows of 60--100, 100--200, and $>$200\GeV, respectively.
For SR1 (SR2--3), we additionally demand $\dPW > 0.7$ ($\dPWH > 0.7$).
Events with two jets, considered as candidates for the resolved radion topology, are categorized into SR4--6 as follows.
The SR4 (SR5) categories have both jets with $60<\MJ<100\GeV$ and require exactly two (one) jets with $\dPW > 0.5$.
Events with $60<\MJmax<100\GeV$ and $\dPW > 0.7$ for the higher-mass jet and $\MJmin<60\GeV$ for the lower-mass jet are placed in SR6.
Requiring $\dWorWH>0.7$ results in a background rejection of approximately 74 (67)\%, while maintaining a signal selection efficiency of about 65 (70)\%.

The \dPW and \dPWH variables are both calibrated in the same data regions enriched in SM \Wjets and top quark events.
To serve as proxies, the SM events are split into various \PW, \Pq/\Pg, and top quark categories mimicking the signal decay structure. Both signal and proxy jets are categorized by geometrically matching parton-level information to the reconstructed jets.
Jets from single \PW boson decays in SM events are used as proxy jets for resolved signal and merged \Rlqq events.
As there is no direct correspondence to any SM event topology for \Rfourq and \Rthreeq events, fully merged top quark jets ($\PQt \to \Pb\Pq\Pq$) serve as proxies in this case.
By performing a simultaneous fit of the proxy templates to the data in regions with different relative compositions, we derive corresponding scale factors (SFs) and associated uncertainties.
These SFs are applied per matched jet category to correct selection efficiencies in simulation for the \dPW and \dPWH spectra.
This calibration procedure is validated in various jet samples.
The detailed procedure is presented in Ref.~\cite{B2G-21-002}.

The main backgrounds, \Wjets and \ttbar production, are estimated using CRs.
The \ttbar CRs are defined by inverting the \PQb jet veto, removing the \dWorWH discriminant selection criteria defined for the SRs, and allowing for up to four additional AK4 jets to increase the number of selected events.
Similarly, for the \Wjets CRs, the \dWorWH selection criteria are inverted, and \ttbar events are vetoed by requiring $\text{deep-}\PQt < 0.4$.
All other backgrounds are estimated using simulation and are subtracted from the data for this procedure.
A linear fit is performed to the ratio of the data to the background of interest (\Wjets or \ttbar), using the \MJlv or \MJJlv distributions, depending on the region, to extract a correction function for the background shape and normalization in the corresponding SR.

The final signal and background yields are determined simultaneously by performing a maximum likelihood fit to the \MJlv and \MJJlv distributions in data for SR1--3 and SR4--6, respectively.
Systematic uncertainties affecting signal and background yields are treated as nuisance parameters and profiled in the statistical interpretation using log-normal and Gaussian constraints for rate and shape uncertainties, respectively.

Uncertainties in the background normalization and shape are derived from the data in the CRs.
In particular, the statistical uncertainty in the CR fits to the \MJlv and \MJJlv distributions is propagated to the SRs through constraints on modeling parameters common to the SR and CR.
Both rate and shape uncertainties are evaluated separately for the \Wjets and top quark backgrounds, and are treated as uncorrelated across the SRs.

Several uncertainties are taken into account for the \textsc{DeepAK8} discriminants and are evaluated as functions of \MJ and \ptj.
Residual differences between data and simulation observed in the validation regions result in a 10\% uncertainty for all jet types.
Additional uncertainties are derived by considering an alternative parton shower simulation and evaluating the effect on the SFs for signal and background jets.
Since the objects used in the calibration procedure have a similar decay structure to the signal, but can exhibit features such as different color flow and quark flavor that affect the \textsc{DeepAK8} performance, additional uncertainties are considered for the signal.
These uncertainties are evaluated based on the shape differences between signal and SM proxy jets in the \dPW and \dPWH spectra.
They amount to 10--40\% for  \Rlqq, \Rthreeq and \Rfourq events, and to 100\% for signal events not matching these categories.
To further account for the different \ptj regimes of signal and proxy jets used in the derivation of the SFs, signal events are simulated with the \HERWIG 2.7 parton shower program~\cite{Bahr:2008pv}.
The resulting differences in the SR yields of up to 25\% are assigned as rate uncertainties.
A detailed description of the uncertainty evaluation procedure can be found in Ref.~\cite{B2G-21-002}.
Uncertainties due to pileup, integrated luminosity, trigger, lepton reconstruction, parton distribution functions (PDFs),
renormalization and factorization scales, and jet energy scale and resolution, largely affecting signal only,
are in total found to be less than 3\% in the rate. They have negligible effect on the shape of the $\Pell\PGn$+jets mass distributions.

The results of this search are statistically combined with those from the search in the fully hadronic final state~\cite{B2G-21-002}.
The SF uncertainties are treated as correlated among the two channels.
Uncertainties in pileup modeling, PDFs, renormalization and factorization scales, as well as the jet energy scale and resolution are also treated as correlated.
All other uncertainties are treated as uncorrelated.

\begin{figure}[htb]
	\centering
	\includegraphics[width=\cmsFigWidth]{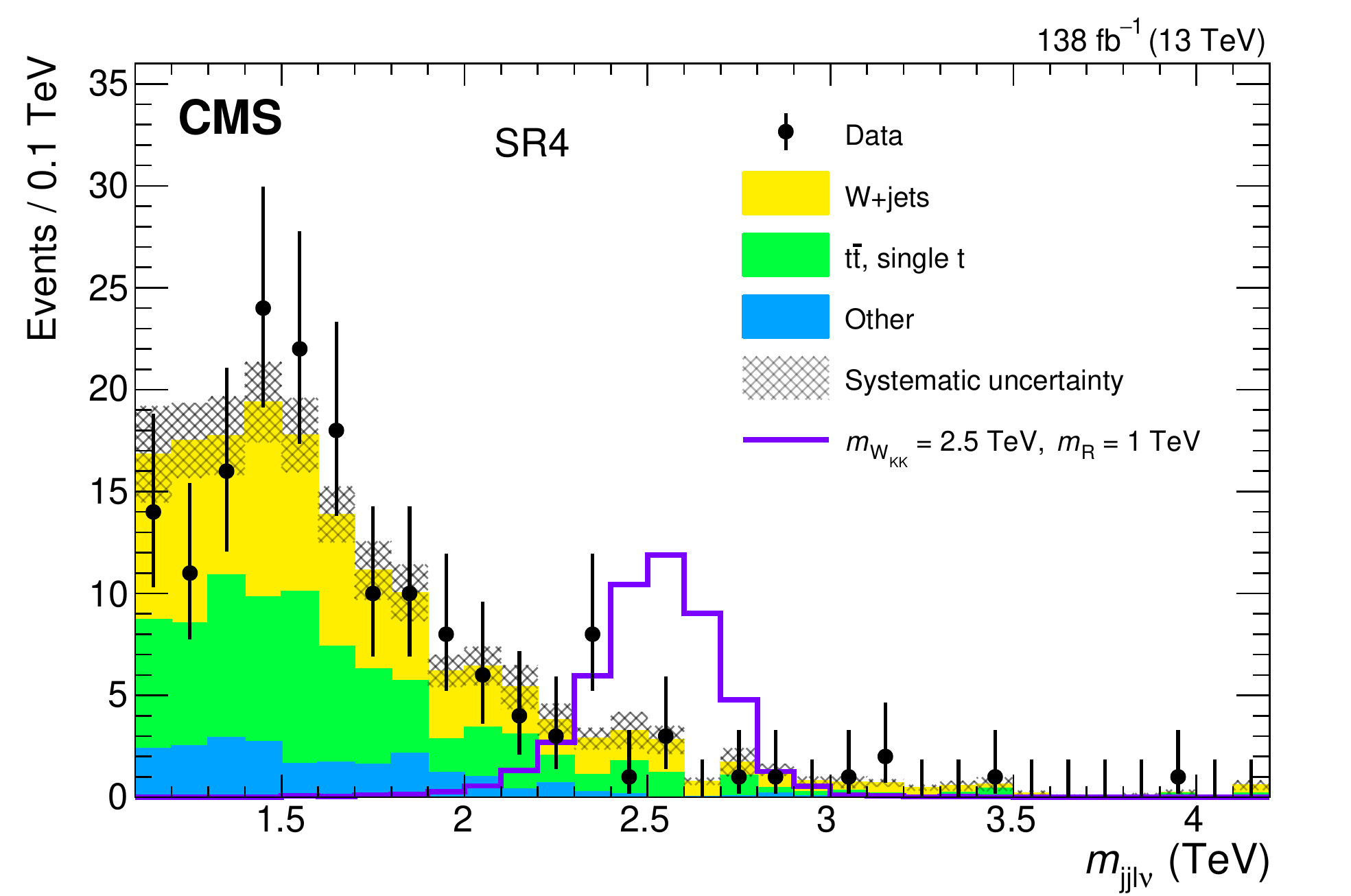}
	\caption{Background-only post-fit distribution of the reconstructed $\Pell\PGn$+jets system \MJJlv in data and simulation for SR4.
	The shape of a triboson signal with $\MWKK = 2.5\TeV$ and $\MR = 1\TeV$ is also shown as a violet solid line,
	normalized to the theoretical production cross section.}
	\label{fig:Mjllnu}
\end{figure}

\begin{figure}[ht!p]
	\centering
	\includegraphics[width=\cmsFigWidth]{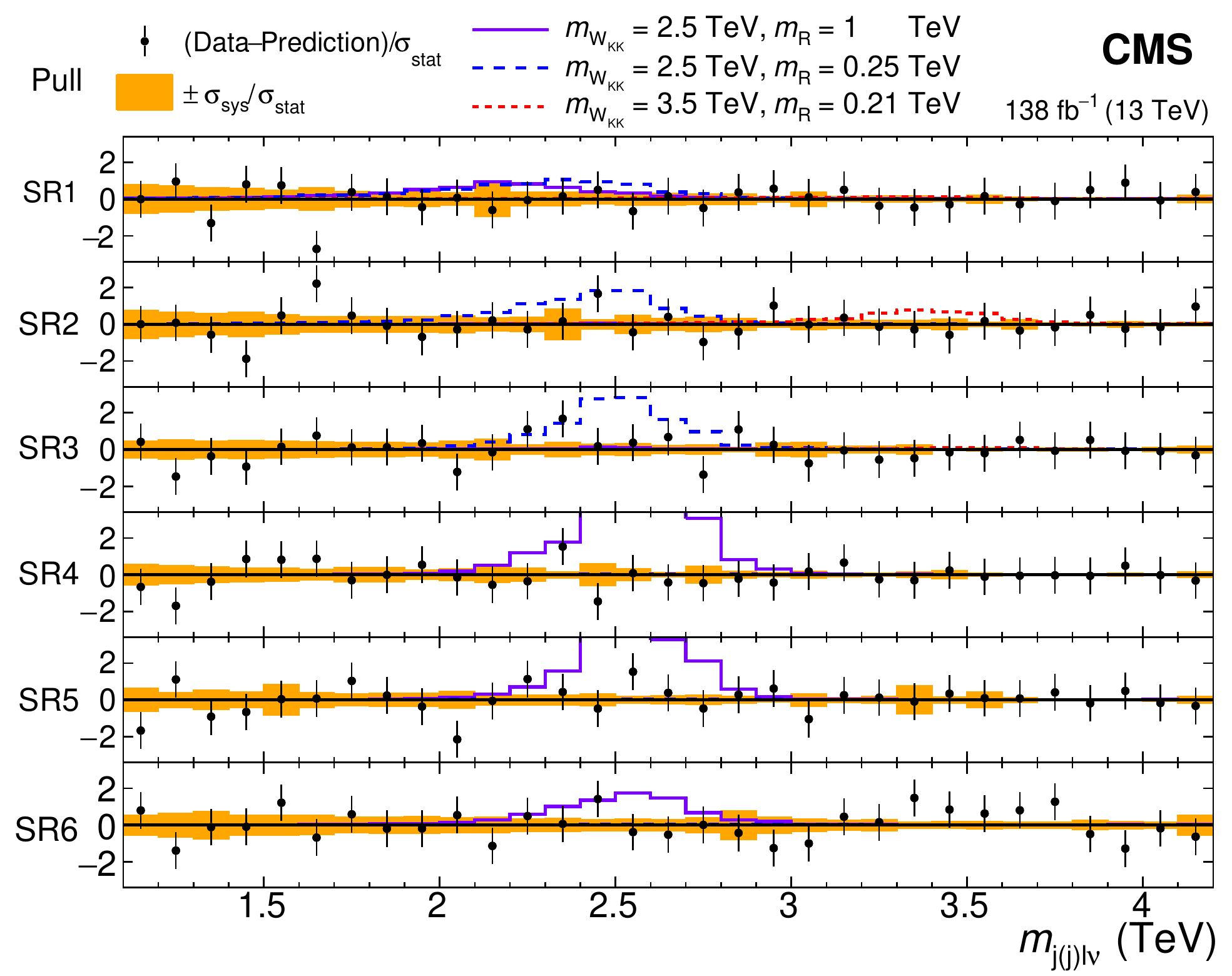}
	\caption{Pull distributions showing (Data$-$Prediction$)/\sigma_{\text{stat}}$ of the background-only fit to the reconstructed $\Pell\PGn$+jets system \MJlv or \MJJlv for all SRs, where $\sigma_{\text{stat}}$ is the statistical uncertainty. 
	Post-fit systematic uncertainties are indicated by the shaded bands.
	Examples of signal scenarios normalized to their theoretical production cross section are shown using solid, dashed, and dotted lines.}
	\label{fig:pulls}
\end{figure}

The background-only post-fit distribution of the reconstructed $\Pell\PGn$+jets system \MJJlv for the most sensitive region for the resolved signal, SR4, is shown in Fig.~\ref{fig:Mjllnu}.
The results for the six SRs of this search are presented in the form of pull distributions [(Data$-$Prediction$)/\sigma_{\text{stat}}$] of the background-only fit in Fig.~\ref{fig:pulls}.
Selected signals have been added on top of the background.
The data are consistent with the background expectation.

\begin{figure}[t!h]
	\centering
	\includegraphics[width=\cmsFigWidth]{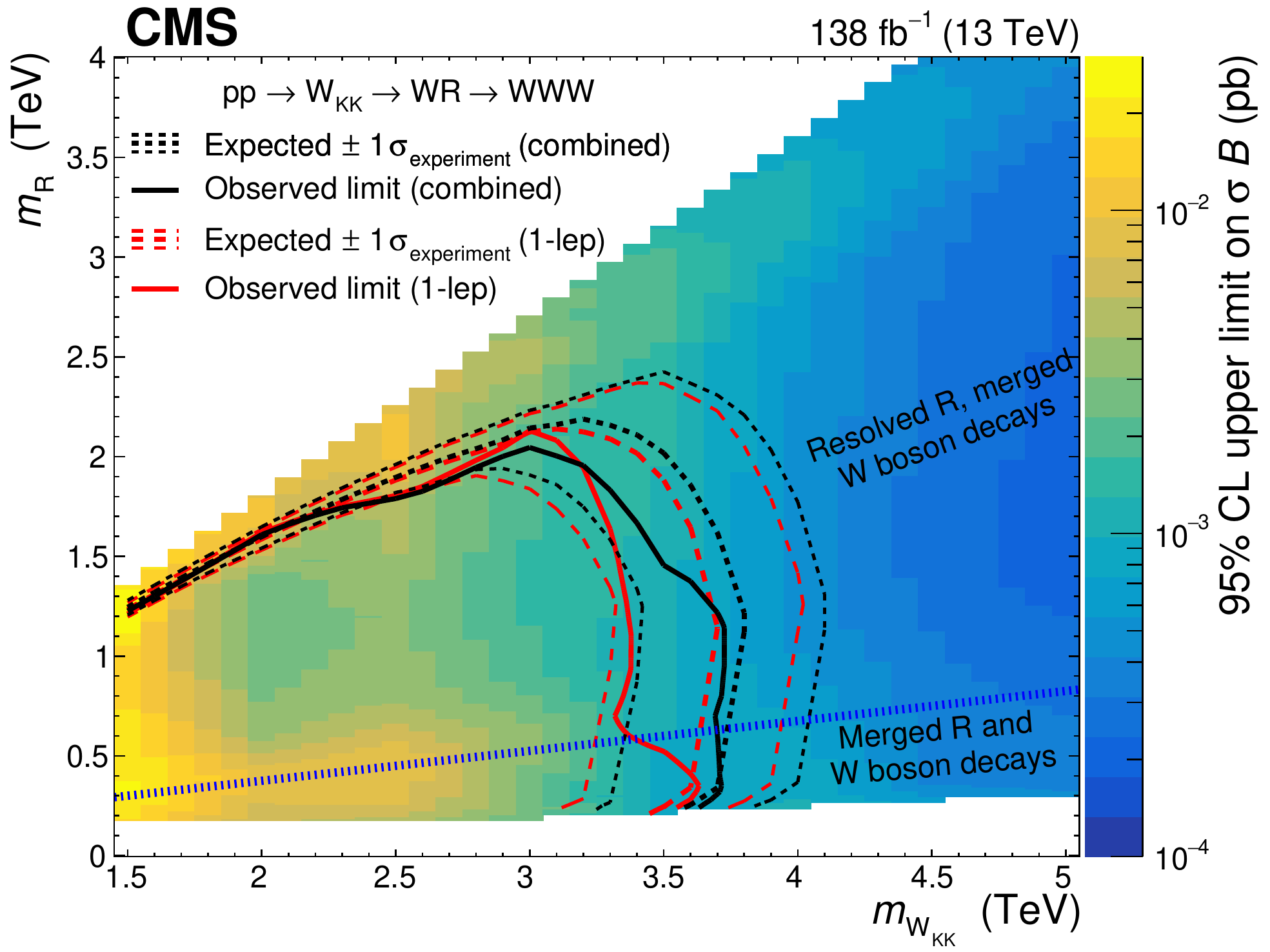}
	\caption{Observed upper limits at 95\% \CL on the product of the signal cross section and the branching fraction ($\sigma \mathcal{B}$) to three \PW bosons as functions of the \WKK and \PR resonance masses.
	Expected (dashed lines) and observed (solid lines) lower mass limits are shown as well for the particular parameters of the explored model.
		The blue straight dashed line indicates the border between merged and resolved radion cases.
		The limits obtained from this analysis are shown in red, and the results of the combination with Ref.~\cite{B2G-21-002} are shown in black.}
	\label{fig:Limits}
\end{figure}

The asymptotic approximation~\cite{Cowan:2010js} of the \CLs technique~\cite{CLS1,CLS2} is used to set limits.
The lower mass limits at 95\% confidence level (\CL) of the $\Pell\PGn$+jets analysis are shown in Fig.~\ref{fig:Limits}.
An excess of events in data around $\MJJlv = 3.5\TeV$ in SR6 results in a weaker than expected observed limit for the resolved signal.
For the combination with the fully hadronic analysis~\cite{B2G-21-002}, lower mass limits are also shown
as well as upper limits on the product of the signal cross section and the branching fraction to three \PW bosons for a resonance with decay width significantly smaller than the detector resolution.
For radion masses between 0.2 and 1.2\TeV, triboson resonances are excluded up to $\MWKK=3.3$ and 3.7\TeV by the $\Pell\PGn$+jets analysis and the combination, respectively.

In summary, a search has been presented for resonances decaying in cascade through $\WKK \to \PW\PR$ and $\PR \to \PW\PW$ to three \PW bosons,
where \WKK is a massive Kaluza--Klein excitation of a gauge boson and \PR is a scalar radion.
The analysis is performed using proton-proton collision data at a center-of-mass energy of 13\TeV, corresponding to an integrated luminosity of 138\fbinv.
The final states considered contain one isolated charged lepton, missing transverse momentum, and one or two massive large-radius jets.
Radion decay configurations with two \PW bosons merged in a single jet and those with two separated \PW boson jets are simultaneously probed by combining jet substructure algorithms.
These novel radion identification and calibration techniques are also applicable to Lorentz-boosted Higgs boson decays.
Results agree with the predictions of the standard model and are combined with those of the analysis in the fully hadronic final state~\cite{B2G-21-002}.
Limits are set on an extended warped extra-dimensional model.
These are the first searches for the production of \TeV-scale triboson resonances at the LHC.

\begin{acknowledgments}
	We congratulate our colleagues in the CERN accelerator departments for the excellent performance of the LHC and thank the technical and administrative staffs at CERN and at other CMS institutes for their contributions to the success of the CMS effort. In addition, we gratefully acknowledge the computing centers and personnel of the Worldwide LHC Computing Grid and other centers for delivering so effectively the computing infrastructure essential to our analyses. Finally, we acknowledge the enduring support for the construction and operation of the LHC, the CMS detector, and the supporting computing infrastructure provided by the following funding agencies: BMBWF and FWF (Austria); FNRS and FWO (Belgium); CNPq, CAPES, FAPERJ, FAPERGS, and FAPESP (Brazil); MES and BNSF (Bulgaria); CERN; CAS, MoST, and NSFC (China); MINCIENCIAS (Colombia); MSES and CSF (Croatia); RIF (Cyprus); SENESCYT (Ecuador); MoER, ERC PUT and ERDF (Estonia); Academy of Finland, MEC, and HIP (Finland); CEA and CNRS/IN2P3 (France); BMBF, DFG, and HGF (Germany); GSRI (Greece); NKFIA (Hungary); DAE and DST (India); IPM (Iran); SFI (Ireland); INFN (Italy); MSIP and NRF (Republic of Korea); MES (Latvia); LAS (Lithuania); MOE and UM (Malaysia); BUAP, CINVESTAV, CONACYT, LNS, SEP, and UASLP-FAI (Mexico); MOS (Montenegro); MBIE (New Zealand); PAEC (Pakistan); MSHE and NSC (Poland); FCT (Portugal); JINR (Dubna); MON, RosAtom, RAS, RFBR, and NRC KI (Russia); MESTD (Serbia); MCIN/AEI and PCTI (Spain); MOSTR (Sri Lanka); Swiss Funding Agencies (Switzerland); MST (Taipei); ThEPCenter, IPST, STAR, and NSTDA (Thailand); TUBITAK and TAEK (Turkey); NASU (Ukraine); STFC (United Kingdom); DOE and NSF (USA).
\end{acknowledgments}

\bibliography{auto_generated}

\providecommand{\href}[2]{#2}\begingroup\raggedright\begin{thebibliography}{10}%
\makeatletter
\providecommand{\hrefCMSnoop }[0]{\@secondoftwo}%
\makeatother
\providecommand{\doi}{\texttt{doi:}\begingroup \urlstyle{tt}\Url}

\bibitem{Aaboud:2018bun}
\hrefCMSnoop {}{{ATLAS Collaboration}, ``{Combination of searches for heavy
  resonances decaying into bosonic and leptonic final states using 36\fbinv of
  proton-proton collision data at $\sqrt{s} = 13\TeV$ with the ATLAS
  detector}'',} \textit{ Phys. Rev. D} \textbf{ 98} (2018) 052008,
  \href{http://dx.doi.org/10.1103/PhysRevD.98.052008}{\doi{10.1103/PhysRevD.98.052008}},
\href{http://www.arXiv.org/abs/1808.02380}{\texttt{arXiv:1808.02380}}.

\bibitem{Sirunyan:2019vgt}
\hrefCMSnoop {}{{CMS Collaboration}, ``{Combination of CMS searches for heavy
  resonances decaying to pairs of bosons or leptons}'',} \textit{ Phys. Lett.
  B} \textbf{ 798} (2019) 134952,
  \href{http://dx.doi.org/10.1016/j.physletb.2019.134952}{\doi{10.1016/j.physletb.2019.134952}},
\href{http://www.arXiv.org/abs/1906.00057}{\texttt{arXiv:1906.00057}}.

\bibitem{ATLAS:2020fry}
\hrefCMSnoop {}{{ATLAS Collaboration}, ``{Search for heavy diboson resonances
  in semileptonic final states in $\Pp\Pp$ collisions at $\sqrt{s}=13$ TeV with
  the ATLAS detector}'',} \textit{ Eur. Phys. J. C} \textbf{ 80} (2020) 1165,
  \href{http://dx.doi.org/10.1140/epjc/s10052-020-08554-y}{\doi{10.1140/epjc/s10052-020-08554-y}},
  \href{http://www.arXiv.org/abs/2004.14636}{\texttt{arXiv:2004.14636}}.

\bibitem{CMS:2021klu}
\hrefCMSnoop {}{{CMS Collaboration}, ``{Search for heavy resonances decaying to
  $\PW\PW$, $\PW\PZ$, or $\PW\PH$ boson pairs in a final state consisting of a
  lepton and a large-radius jet in proton-proton collisions at $\sqrt{s}$ =
  13\TeV}'',} \textit{ Phys. Rev. D} \textbf{ 105} (2022) 032008,
  \href{http://dx.doi.org/10.1103/PhysRevD.105.032008}{\doi{10.1103/PhysRevD.105.032008}},
  \href{http://www.arXiv.org/abs/2109.06055}{\texttt{arXiv:2109.06055}}.

\bibitem{CMS-PAPERS-B2G-18-002}
\hrefCMSnoop {}{{CMS Collaboration}, ``{A multi-dimensional search for new
  heavy resonances decaying to boosted $\PW\PW$, $\PW\PZ$, or $\PZ\PZ$ boson
  pairs in the dijet final state at 13 {TeV}}'',} \textit{ Eur. Phys. J. C}
  \textbf{ 80} (2020) 237,
  \href{http://dx.doi.org/10.1140/epjc/s10052-020-7773-5}{\doi{10.1140/epjc/s10052-020-7773-5}},
  \href{http://www.arXiv.org/abs/1906.05977}{\texttt{arXiv:1906.05977}}.

\bibitem{ATLAS:2019nat}
\hrefCMSnoop {}{{ATLAS Collaboration}, ``{Search for diboson resonances in
  hadronic final states in 139 fb$^{-1}$ of $\Pp\Pp$ collisions at $\sqrt{s} =
  13$ TeV with the ATLAS detector}'',} \textit{ JHEP} \textbf{ 09} (2019) 091,
  \href{http://dx.doi.org/10.1007/JHEP09(2019)091}{\doi{10.1007/JHEP09(2019)091}},
  \href{http://www.arXiv.org/abs/1906.08589}{\texttt{arXiv:1906.08589}}.
  [Erratum: \href{https://doi.org/10.1007/JHEP06(2020)042}{JHEP 06, 042
  (2020)}].

\bibitem{Asquith:2018igt}
R.~Kogler\hrefCMSnoop {}{ {et~al.}, ``{Jet substructure at the Large Hadron
  Collider}'',} \textit{ Rev. Mod. Phys.} \textbf{ 91} (2019) 045003,
  \href{http://dx.doi.org/10.1103/RevModPhys.91.045003}{\doi{10.1103/RevModPhys.91.045003}},
  \href{http://www.arXiv.org/abs/1803.06991}{\texttt{arXiv:1803.06991}}.

\bibitem{ATLAS:2020gxx}
\hrefCMSnoop {}{{ATLAS Collaboration}, ``{Search for a heavy Higgs boson
  decaying into a \PZ boson and another heavy Higgs boson in the $\ell \ell bb$
  and $\ell \ell \PW\PW$ final states in $\Pp\Pp$ collisions at $\sqrt{s}=13$
  $\text {TeV}$ with the ATLAS detector}'',} \textit{ Eur. Phys. J. C} \textbf{
  81} (2021) 396,
  \href{http://dx.doi.org/10.1140/epjc/s10052-021-09117-5}{\doi{10.1140/epjc/s10052-021-09117-5}},
  \href{http://www.arXiv.org/abs/2011.05639}{\texttt{arXiv:2011.05639}}.

\bibitem{Aguilar-Saavedra:2015iew}
\hrefCMSnoop {}{J.~A. Aguilar-Saavedra and F.~R. Joaquim, ``{Multiboson
  production in $\PW'$ decays}'',} \textit{ JHEP} \textbf{ 01} (2016) 183,
  \href{http://dx.doi.org/10.1007/JHEP01(2016)183}{\doi{10.1007/JHEP01(2016)183}},
  \href{http://www.arXiv.org/abs/1512.00396}{\texttt{arXiv:1512.00396}}.

\bibitem{Aguilar-Saavedra:2015rna}
\hrefCMSnoop {}{J.~A. Aguilar-Saavedra, ``{Triboson interpretations of the
  ATLAS diboson excess}'',} \textit{ JHEP} \textbf{ 10} (2015) 099,
  \href{http://dx.doi.org/10.1007/JHEP10(2015)099}{\doi{10.1007/JHEP10(2015)099}},
  \href{http://www.arXiv.org/abs/1506.06739}{\texttt{arXiv:1506.06739}}.

\bibitem{Agashe:2016rle}
\hrefCMSnoop {}{K.~Agashe, P.~Du, S.~Hong, and R.~Sundrum, ``{Flavor universal
  resonances and warped gravity}'',} \textit{ JHEP} \textbf{ 01} (2017) 016,
  \href{http://dx.doi.org/10.1007/JHEP01(2017)016}{\doi{10.1007/JHEP01(2017)016}},
  \href{http://www.arXiv.org/abs/1608.00526}{\texttt{arXiv:1608.00526}}.

\bibitem{Agashe:2016kfr}
K.~Agashe\hrefCMSnoop {}{ {et~al.}, ``{LHC signals from cascade decays of
  warped vector resonances}'',} \textit{ JHEP} \textbf{ 05} (2017) 078,
  \href{http://dx.doi.org/10.1007/JHEP05(2017)078}{\doi{10.1007/JHEP05(2017)078}},
  \href{http://www.arXiv.org/abs/1612.00047}{\texttt{arXiv:1612.00047}}.

\bibitem{Agashe:2017wss}
K.~Agashe\hrefCMSnoop {}{ {et~al.}, ``{Dedicated strategies for triboson
  signals from cascade decays of vector resonances}'',} \textit{ Phys. Rev. D}
  \textbf{ 99} (2019) 075016,
  \href{http://dx.doi.org/10.1103/PhysRevD.99.075016}{\doi{10.1103/PhysRevD.99.075016}},
  \href{http://www.arXiv.org/abs/1711.09920}{\texttt{arXiv:1711.09920}}.

\bibitem{Aguilar-Saavedra:2017iso}
\hrefCMSnoop {}{J.~A. Aguilar-Saavedra, ``{Profile of multiboson signals}'',}
  \textit{ JHEP} \textbf{ 05} (2017) 066,
  \href{http://dx.doi.org/10.1007/JHEP05(2017)066}{\doi{10.1007/JHEP05(2017)066}},
  \href{http://www.arXiv.org/abs/1703.06153}{\texttt{arXiv:1703.06153}}.

\bibitem{Agashe:2018leo}
K.~Agashe\hrefCMSnoop {}{ {et~al.}, ``{Detecting a boosted diboson
  resonance}'',} \textit{ JHEP} \textbf{ 11} (2018) 027,
  \href{http://dx.doi.org/10.1007/JHEP11(2018)027}{\doi{10.1007/JHEP11(2018)027}},
  \href{http://www.arXiv.org/abs/1809.07334}{\texttt{arXiv:1809.07334}}.

\bibitem{Kuang:2015gka}
\hrefCMSnoop {}{Y.-P. Kuang, H.-Y. Ren, and L.-H. Xia, ``{Further investigation
  of the model-independent probe of heavy neutral Higgs bosons at LHC Run
  2}'',} \textit{ Chin. Phys. C} \textbf{ 40} (2016) 023101,
  \href{http://dx.doi.org/10.1088/1674-1137/40/2/023101}{\doi{10.1088/1674-1137/40/2/023101}},
  \href{http://www.arXiv.org/abs/1506.08007}{\texttt{arXiv:1506.08007}}.

\bibitem{Ren:2014bza}
\hrefCMSnoop {}{Y.-P. Kuang, H.-Y. Ren, and L.-H. Xia, ``{Model-independent
  probe of anomalous heavy neutral Higgs bosons at the LHC}'',} \textit{ Phys.
  Rev. D} \textbf{ 90} (2014) 115002,
  \href{http://dx.doi.org/10.1103/PhysRevD.90.115002}{\doi{10.1103/PhysRevD.90.115002}},
  \href{http://www.arXiv.org/abs/1404.6367}{\texttt{arXiv:1404.6367}}.

\bibitem{B2G-21-002}
\hrefCMSnoop {}{{CMS Collaboration}, ``{Search for resonances decaying to three
  \PW bosons in the hadronic final state in proton-proton collisions at
  $\sqrt{s}$ = 13\TeV}'',} 2021.
  \href{http://www.arXiv.org/abs/2112.13090}{\texttt{arXiv:2112.13090}}.
  Submitted to \textit{Phys. Rev. D.}

\bibitem{Randall:1999ee}
\hrefCMSnoop {}{L.~Randall and R.~Sundrum, ``A large mass hierarchy from a
  small extra dimension'',} \textit{ Phys. Rev. Lett.} \textbf{ 83} (1999)
  3370,
  \href{http://dx.doi.org/10.1103/PhysRevLett.83.3370}{\doi{10.1103/PhysRevLett.83.3370}},
\href{http://www.arXiv.org/abs/hep-ph/9905221}{\texttt{arXiv:hep-ph/9905221}}.

\bibitem{Randall:1999vf}
\hrefCMSnoop {}{L.~Randall and R.~Sundrum, ``An alternative to
  compactification'',} \textit{ Phys. Rev. Lett.} \textbf{ 83} (1999) 4690,
  \href{http://dx.doi.org/10.1103/PhysRevLett.83.4690}{\doi{10.1103/PhysRevLett.83.4690}},
\href{http://www.arXiv.org/abs/hep-th/9906064}{\texttt{arXiv:hep-th/9906064}}.

\bibitem{Agashe:2007zd}
\hrefCMSnoop {}{K.~Agashe, H.~Davoudiasl, G.~Perez, and A.~Soni, ``Warped
  gravitons at the {LHC} and beyond'',} \textit{ Phys. Rev. D} \textbf{ 76}
  (2007) 036006,
  \href{http://dx.doi.org/10.1103/PhysRevD.76.036006}{\doi{10.1103/PhysRevD.76.036006}},
\href{http://www.arXiv.org/abs/hep-ph/0701186}{\texttt{arXiv:hep-ph/0701186}}.

\bibitem{Fitzpatrick:2007qr}
\hrefCMSnoop {}{A.~L. Fitzpatrick, J.~Kaplan, L.~Randall, and L.-T. Wang,
  ``Searching for the {Kaluza}-{Klein} graviton in bulk {RS} models'',}
  \textit{ JHEP} \textbf{ 09} (2007) 013,
  \href{http://dx.doi.org/10.1088/1126-6708/2007/09/013}{\doi{10.1088/1126-6708/2007/09/013}},
\href{http://www.arXiv.org/abs/hep-ph/0701150}{\texttt{arXiv:hep-ph/0701150}}.

\bibitem{Goldberger_1999}
\hrefCMSnoop {}{W.~D. Goldberger and M.~B. Wise, ``Modulus stabilization with
  bulk fields'',} \textit{ Phys. Rev. Lett.} \textbf{ 83} (1999) 4922,
  \href{http://dx.doi.org/10.1103/physrevlett.83.4922}{\doi{10.1103/physrevlett.83.4922}},
\href{http://www.arXiv.org/abs/hep-ph/9907447}{\texttt{arXiv:hep-ph/9907447}}.

\bibitem{hepdata}
\hrefCMSnoop {}{``{HEPD}ata record for this analysis'',} 2021.
\newblock
  \href{http://dx.doi.org/10.17182/hepdata.102646}{\doi{10.17182/hepdata.102646}}.

\bibitem{CMS-LUM-17-003}
\hrefCMSnoop {}{{CMS Collaboration}, ``{Precision luminosity measurement in
  proton-proton collisions at $\sqrt{s} =$ 13 TeV in 2015 and 2016 at CMS}'',}
  \textit{ Eur. Phys. J. C} \textbf{ 81} (2021) 800,
  \href{http://dx.doi.org/10.1140/epjc/s10052-021-09538-2}{\doi{10.1140/epjc/s10052-021-09538-2}},
\href{http://www.arXiv.org/abs/2104.01927}{\texttt{arXiv:2104.01927}}.

\bibitem{CMS-PAS-LUM-17-004}
\href {https://cds.cern.ch/record/2621960}{{CMS Collaboration}, ``{CMS
  luminosity measurement for the 2017 data-taking period at $\sqrt{s} =
  13~\mathrm{TeV}$}'',} CMS Physics Analysis Summary CMS-PAS-LUM-17-004, 2018.

\bibitem{CMS-PAS-LUM-18-002}
\href {https://cds.cern.ch/record/2676164}{{CMS Collaboration}, ``{CMS
  luminosity measurement for the 2018 data-taking period at $\sqrt{s} =
  13~\mathrm{TeV}$}'',} CMS Physics Analysis Summary CMS-PAS-LUM-18-002, 2019.

\bibitem{CMS:2008xjf}
\hrefCMSnoop {}{{CMS Collaboration}, ``The {CMS} experiment at the {CERN}
  {LHC}'',} \textit{ JINST} \textbf{ 3} (2008) S08004,
  \href{http://dx.doi.org/10.1088/1748-0221/3/08/S08004}{\doi{10.1088/1748-0221/3/08/S08004}}.

\bibitem{CMS:2020cmk}
\hrefCMSnoop {}{{CMS Collaboration}, ``{Performance of the CMS Level-1 trigger
  in proton-proton collisions at $\sqrt{s} = 13$\,TeV}'',} \textit{ JINST}
  \textbf{ 15} (2020) P10017,
  \href{http://dx.doi.org/10.1088/1748-0221/15/10/P10017}{\doi{10.1088/1748-0221/15/10/P10017}},
  \href{http://www.arXiv.org/abs/2006.10165}{\texttt{arXiv:2006.10165}}.

\bibitem{CMS:2016ngn}
\hrefCMSnoop {}{{CMS Collaboration}, ``{The CMS trigger system}'',} \textit{
  JINST} \textbf{ 12} (2017) P01020,
  \href{http://dx.doi.org/10.1088/1748-0221/12/01/P01020}{\doi{10.1088/1748-0221/12/01/P01020}},
\href{http://www.arXiv.org/abs/1609.02366}{\texttt{arXiv:1609.02366}}.

\bibitem{CMS:2015xaf}
\hrefCMSnoop {}{{CMS Collaboration}, ``{Performance of electron reconstruction
  and selection with the CMS detector in proton-proton collisions at $\sqrt{s}
  = 8$\TeV}'',} \textit{ JINST} \textbf{ 10} (2015) P06005,
  \href{http://dx.doi.org/10.1088/1748-0221/10/06/P06005}{\doi{10.1088/1748-0221/10/06/P06005}},
\href{http://www.arXiv.org/abs/1502.02701}{\texttt{arXiv:1502.02701}}.

\bibitem{CMS:2018rym}
\hrefCMSnoop {}{{CMS Collaboration}, ``{Performance of the CMS muon detector
  and muon reconstruction with proton-proton collisions at $\sqrt{s}=$ 13
  TeV}'',} \textit{ JINST} \textbf{ 13} (2018) P06015,
  \href{http://dx.doi.org/10.1088/1748-0221/13/06/P06015}{\doi{10.1088/1748-0221/13/06/P06015}},
  \href{http://www.arXiv.org/abs/1804.04528}{\texttt{arXiv:1804.04528}}.

\bibitem{CMS:2015myp}
\hrefCMSnoop {}{{CMS Collaboration}, ``Performance of photon reconstruction and
  identification with the {CMS} detector in proton-proton collisions at
  $\sqrt{s}$ = 8 {TeV}'',} \textit{ JINST} \textbf{ 10} (2015) P08010,
  \href{http://dx.doi.org/10.1088/1748-0221/10/08/P08010}{\doi{10.1088/1748-0221/10/08/P08010}},
  \href{http://www.arXiv.org/abs/1502.02702}{\texttt{arXiv:1502.02702}}.

\bibitem{CMS:2014pgm}
\hrefCMSnoop {}{{CMS Collaboration}, ``{Description and performance of track
  and primary-vertex reconstruction with the CMS tracker}'',} \textit{ JINST}
  \textbf{ 9} (2014) P10009,
  \href{http://dx.doi.org/10.1088/1748-0221/9/10/P10009}{\doi{10.1088/1748-0221/9/10/P10009}},
  \href{http://www.arXiv.org/abs/1405.6569}{\texttt{arXiv:1405.6569}}.

\bibitem{CMS:2017yfk}
\hrefCMSnoop {}{{CMS Collaboration}, ``Particle-flow reconstruction and global
  event description with the {CMS} detector'',} \textit{ JINST} \textbf{ 12}
  (2017) P10003,
  \href{http://dx.doi.org/10.1088/1748-0221/12/10/P10003}{\doi{10.1088/1748-0221/12/10/P10003}},
\href{http://www.arXiv.org/abs/1706.04965}{\texttt{arXiv:1706.04965}}.

\bibitem{CMS:2018jrd}
\hrefCMSnoop {}{{CMS Collaboration}, ``{Performance of reconstruction and
  identification of $\tau$ leptons decaying to hadrons and $\nu_\tau$ in
  $\Pp\Pp$ collisions at $\sqrt{s}=$ 13 TeV}'',} \textit{ JINST} \textbf{ 13}
  (2018) P10005,
  \href{http://dx.doi.org/10.1088/1748-0221/13/10/P10005}{\doi{10.1088/1748-0221/13/10/P10005}},
  \href{http://www.arXiv.org/abs/1809.02816}{\texttt{arXiv:1809.02816}}.

\bibitem{CMS:2016lmd}
\hrefCMSnoop {}{{CMS Collaboration}, ``{Jet energy scale and resolution in the
  CMS experiment in $\Pp\Pp$ collisions at 8 TeV}'',} \textit{ JINST} \textbf{
  12} (2017) P02014,
  \href{http://dx.doi.org/10.1088/1748-0221/12/02/P02014}{\doi{10.1088/1748-0221/12/02/P02014}},
\href{http://www.arXiv.org/abs/1607.03663}{\texttt{arXiv:1607.03663}}.

\bibitem{CMS:2019ctu}
\hrefCMSnoop {}{{CMS Collaboration}, ``Performance of missing transverse
  momentum reconstruction in proton-proton collisions at $\sqrt{s} = 13$\,{TeV}
  using the {CMS} detector'',} \textit{ JINST} \textbf{ 14} (2019) P07004,
  \href{http://dx.doi.org/10.1088/1748-0221/14/07/P07004}{\doi{10.1088/1748-0221/14/07/P07004}},
\href{http://www.arXiv.org/abs/1903.06078}{\texttt{arXiv:1903.06078}}.

\bibitem{Alwall:2014hca}
J.~Alwall\hrefCMSnoop {}{ {et~al.}, ``{The automated computation of tree-level
  and next-to-leading order differential cross sections, and their matching to
  parton shower simulations}'',} \textit{ JHEP} \textbf{ 07} (2014) 079,
  \href{http://dx.doi.org/10.1007/JHEP07(2014)079}{\doi{10.1007/JHEP07(2014)079}},
\href{http://www.arXiv.org/abs/1405.0301}{\texttt{arXiv:1405.0301}}.

\bibitem{Alioli:2011as}
\hrefCMSnoop {}{S.~Alioli, S.-O. Moch, and P.~Uwer, ``{Hadronic top-quark
  pair-production with one jet and parton showering}'',} \textit{ JHEP}
  \textbf{ 01} (2012) 137,
  \href{http://dx.doi.org/10.1007/JHEP01(2012)137}{\doi{10.1007/JHEP01(2012)137}},
\href{http://www.arXiv.org/abs/1110.5251}{\texttt{arXiv:1110.5251}}.

\bibitem{Sjostrand:2014zea}
T.~Sj{\"o}strand\hrefCMSnoop {}{ {et~al.}, ``{An introduction to PYTHIA
  8.2}'',} \textit{ Comput. Phys. Commun.} \textbf{ 191} (2015) 159,
  \href{http://dx.doi.org/10.1016/j.cpc.2015.01.024}{\doi{10.1016/j.cpc.2015.01.024}},
\href{http://www.arXiv.org/abs/1410.3012}{\texttt{arXiv:1410.3012}}.

\bibitem{Agostinelli:2002hh}
\hrefCMSnoop {}{{GEANT4} Collaboration, ``{\GEANTfour}---a simulation
  toolkit'',} \textit{ Nucl. Instrum. Meth. A} \textbf{ 506} (2003) 250,
\href{http://dx.doi.org/10.1016/S0168-9002(03)01368-8}{\doi{10.1016/S0168-9002(03)01368-8}}.

\bibitem{Sirunyan:2020foa}
\hrefCMSnoop {}{{CMS Collaboration}, ``{Pileup mitigation at CMS in 13 TeV
  data}'',} \textit{ JINST} \textbf{ 15} (2020) P09018,
  \href{http://dx.doi.org/10.1088/1748-0221/15/09/P09018}{\doi{10.1088/1748-0221/15/09/P09018}},
  \href{http://www.arXiv.org/abs/2003.00503}{\texttt{arXiv:2003.00503}}.

\bibitem{Cacciari:2008gp}
\hrefCMSnoop {}{M.~Cacciari, G.~P. Salam, and G.~Soyez, ``The anti-\kt jet
  clustering algorithm'',} \textit{ JHEP} \textbf{ 04} (2008) 063,
  \href{http://dx.doi.org/10.1088/1126-6708/2008/04/063}{\doi{10.1088/1126-6708/2008/04/063}},
  \href{http://www.arXiv.org/abs/0802.1189}{\texttt{arXiv:0802.1189}}.

\bibitem{Cacciari:2011ma}
\hrefCMSnoop {}{M.~Cacciari, G.~P. Salam, and G.~Soyez, ``{FastJet user
  manual}'',} \textit{ Eur. Phys. J. C} \textbf{ 72} (2012) 1896,
  \href{http://dx.doi.org/10.1140/epjc/s10052-012-1896-2}{\doi{10.1140/epjc/s10052-012-1896-2}},
\href{http://www.arXiv.org/abs/1111.6097}{\texttt{arXiv:1111.6097}}.

\bibitem{Sirunyan:2017ezt}
\hrefCMSnoop {}{{CMS Collaboration}, ``{Identification of heavy-flavour jets
  with the CMS detector in $\Pp\Pp$ collisions at 13 TeV}'',} \textit{ JINST}
  \textbf{ 13} (2018) P05011,
  \href{http://dx.doi.org/10.1088/1748-0221/13/05/P05011}{\doi{10.1088/1748-0221/13/05/P05011}},
\href{http://www.arXiv.org/abs/1712.07158}{\texttt{arXiv:1712.07158}}.

\bibitem{Dasgupta:2013ihk}
\hrefCMSnoop {}{M.~Dasgupta, A.~Fregoso, S.~Marzani, and G.~P. Salam, ``Towards
  an understanding of jet substructure'',} \textit{ JHEP} \textbf{ 09} (2013)
  029,
  \href{http://dx.doi.org/10.1007/JHEP09(2013)029}{\doi{10.1007/JHEP09(2013)029}},
\href{http://www.arXiv.org/abs/1307.0007}{\texttt{arXiv:1307.0007}}.

\bibitem{Butterworth:2008iy}
\hrefCMSnoop {}{J.~M. Butterworth, A.~R. Davison, M.~Rubin, and G.~P. Salam,
  ``Jet substructure as a new {Higgs} search channel at the {LHC}'',} \textit{
  Phys. Rev. Lett.} \textbf{ 100} (2008) 242001,
  \href{http://dx.doi.org/10.1103/PhysRevLett.100.242001}{\doi{10.1103/PhysRevLett.100.242001}},
\href{http://www.arXiv.org/abs/0802.2470}{\texttt{arXiv:0802.2470}}.

\bibitem{Larkoski:2014wba}
\hrefCMSnoop {}{A.~J. Larkoski, S.~Marzani, G.~Soyez, and J.~Thaler, ``Soft
  drop'',} \textit{ JHEP} \textbf{ 05} (2014) 146,
  \href{http://dx.doi.org/10.1007/JHEP05(2014)146}{\doi{10.1007/JHEP05(2014)146}},
\href{http://www.arXiv.org/abs/1402.2657}{\texttt{arXiv:1402.2657}}.

\bibitem{Sirunyan:2018iff}
\hrefCMSnoop {}{{CMS Collaboration}, ``{Search for a heavy resonance decaying
  to a pair of vector bosons in the lepton plus merged jet final state at $
  \sqrt{s}=13 $ TeV}'',} \textit{ JHEP} \textbf{ 05} (2018) 088,
  \href{http://dx.doi.org/10.1007/JHEP05(2018)088}{\doi{10.1007/JHEP05(2018)088}},
  \href{http://www.arXiv.org/abs/1802.09407}{\texttt{arXiv:1802.09407}}.

\bibitem{Sirunyan:2020lcu}
\hrefCMSnoop {}{{CMS Collaboration}, ``{Identification of heavy, energetic,
  hadronically decaying particles using machine-learning techniques}'',}
  \textit{ JINST} \textbf{ 15} (2020) P06005,
  \href{http://dx.doi.org/10.1088/1748-0221/15/06/P06005}{\doi{10.1088/1748-0221/15/06/P06005}},
  \href{http://www.arXiv.org/abs/2004.08262}{\texttt{arXiv:2004.08262}}.

\bibitem{Bahr:2008pv}
\hrefCMSnoop {}{M.~Bahr {et~al.}, ``{Herwig++ physics and manual}'',} \textit{
  Eur. Phys. J. C} \textbf{ 58} (2008) 639,
  \href{http://dx.doi.org/10.1140/epjc/s10052-008-0798-9}{\doi{10.1140/epjc/s10052-008-0798-9}},
  \href{http://www.arXiv.org/abs/0803.0883}{\texttt{arXiv:0803.0883}}.

\bibitem{Cowan:2010js}
\hrefCMSnoop {}{G.~Cowan, K.~Cranmer, E.~Gross, and O.~Vitells, ``Asymptotic
  formulae for likelihood-based tests of new physics'',} \textit{ Eur. Phys. J.
  C} \textbf{ 71} (2011) 1554,
  \href{http://dx.doi.org/10.1140/epjc/s10052-011-1554-0}{\doi{10.1140/epjc/s10052-011-1554-0}},
  \href{http://www.arXiv.org/abs/1007.1727}{\texttt{arXiv:1007.1727}}.
[Erratum: \href{https://doi.org/10.1140/epjc/s10052-013-2501-z}{Eur. Phys. J. C
  \textbf{73}, 2501 (2013)}].

\bibitem{CLS1}
\hrefCMSnoop {}{T.~Junk, ``Confidence level computation for combining searches
  with small statistics'',} \textit{ Nucl. Instrum. Meth. A} \textbf{ 434}
  (1999) 435,
  \href{http://dx.doi.org/10.1016/S0168-9002(99)00498-2}{\doi{10.1016/S0168-9002(99)00498-2}},
\href{http://www.arXiv.org/abs/hep-ex/9902006}{\texttt{arXiv:hep-ex/9902006}}.

\bibitem{CLS2}
\hrefCMSnoop {}{A.~L. Read, ``Presentation of search results: The
  {CL$_{\text{s}}$} technique'',} \textit{ J. Phys. G} \textbf{ 28} (2002)
  2693,
\href{http://dx.doi.org/10.1088/0954-3899/28/10/313}{\doi{10.1088/0954-3899/28/10/313}}.

\end{thebibliography}\endgroup
\cleardoublepage \appendix\section{The CMS Collaboration \label{app:collab}}\begin{sloppypar}\hyphenpenalty=5000\widowpenalty=500\clubpenalty=5000\cmsinstitute{Yerevan~Physics~Institute, Yerevan, Armenia}
A.~Tumasyan
\cmsinstitute{Institut~f\"{u}r~Hochenergiephysik, Vienna, Austria}
W.~Adam\cmsorcid{0000-0001-9099-4341}, J.W.~Andrejkovic, T.~Bergauer\cmsorcid{0000-0002-5786-0293}, S.~Chatterjee\cmsorcid{0000-0003-2660-0349}, M.~Dragicevic\cmsorcid{0000-0003-1967-6783}, A.~Escalante~Del~Valle\cmsorcid{0000-0002-9702-6359}, R.~Fr\"{u}hwirth\cmsAuthorMark{1}, M.~Jeitler\cmsAuthorMark{1}\cmsorcid{0000-0002-5141-9560}, N.~Krammer, L.~Lechner\cmsorcid{0000-0002-3065-1141}, D.~Liko, I.~Mikulec, P.~Paulitsch, F.M.~Pitters, J.~Schieck\cmsAuthorMark{1}\cmsorcid{0000-0002-1058-8093}, R.~Sch\"{o}fbeck\cmsorcid{0000-0002-2332-8784}, M.~Spanring\cmsorcid{0000-0001-6328-7887}, S.~Templ\cmsorcid{0000-0003-3137-5692}, W.~Waltenberger\cmsorcid{0000-0002-6215-7228}, C.-E.~Wulz\cmsAuthorMark{1}\cmsorcid{0000-0001-9226-5812}
% ...
\vskip\cmsinstskip
% ...
1:~Also at TU Wien, Wien, Austria\\
% ...
\end{sloppypar}
\end{document}